\documentclass[aps,noshowpacs,floatfix,superscriptaddress,preprintnumbers,showpacs,showkeys,nofootinbib]{revtex4}
\usepackage{graphicx}
\usepackage{amsmath,amssymb}
\usepackage{subfigure}
\usepackage{slashed,braket}
\usepackage{bbold}

\newcommand{\be}{\begin{equation}}
\newcommand{\ee}{\end{equation}}
\newcommand{\ba}{\begin{eqnarray}}
\newcommand{\ea}{\end{eqnarray}}

\newcommand{\eq}{\begin{eqnarray}}
\newcommand{\en}{\end{eqnarray}}

\begin{document}

\title{Studying the $P_c(4450)$ resonance in $J/\psi$ photoproduction off protons
}


\author{Astrid N.~Hiller Blin}
\affiliation{Institut f\"ur Kernphysik \& PRISMA Cluster of Excellence, Johannes Gutenberg Universit\"at, D-55099 Mainz, Germany}
\author{C\'esar Fern\'andez-Ram\'irez}
\affiliation{Instituto de Ciencias Nucleares, Universidad Nacional Aut\'onoma de M\'exico, Ciudad de M\'exico 04510, Mexico}
\author{Andrew Jackura}
\affiliation{Center for Exploration of Energy and Matter, Indiana University, Bloomington, IN 47403, USA}
\affiliation{Physics Department, Indiana University, Bloomington, IN 47405, USA}
\author{Vincent Mathieu}
\affiliation{Thomas Jefferson National Accelerator Facility, Newport News, VA 23606, USA}
\author{Viktor~I.~Mokeev}
\affiliation{Thomas Jefferson National Accelerator Facility, Newport News, VA 23606, USA}
\author{Alessandro Pilloni}
\affiliation{Thomas Jefferson National Accelerator Facility, Newport News, VA 23606, USA}
\author{Adam P.~Szczepaniak}
\affiliation{Center for Exploration of Energy and Matter, Indiana University, Bloomington, IN 47403, USA}
\affiliation{Physics Department, Indiana University, Bloomington, IN 47405, USA}
\affiliation{Thomas Jefferson National Accelerator Facility, Newport News, VA 23606, USA}

\today

\begin{abstract}
The LHCb has reported the observation of a resonancelike structure, the $P_c(4450)$, in the $J/\psi~p$ invariant masses. In our work, we discuss the feasibility of detecting this structure in $J/\psi$ photoproduction, e.g.\ in the measurements that have been approved for the experiments in Hall A/C and in Hall B with CLAS12 at JLab. Also the GlueX Collaboration has already reported preliminary results. 
We take into account the experimental resolution effects, and perform a global fit to world $J/\psi$ photoproduction data in order to study the possibility of observing the $P_c(4450)$ signal in future JLab data. We present a first estimate of the upper limit for the branching ratio of the $P_c(4450)$ into the $J/\psi~p$ channel, and we study the angular distributions of the differential cross sections. This will shed light on the nature and couplings of the $P_c(4450)$ structure in the future photoproduction experiments.
\end{abstract}

\maketitle

\section{Introduction}
In 2015, the LHCb collaboration announced two resonancelike structures in the $\Lambda^0_b \to K^- (J/\psi~p)$ channel~\cite{Aaij:2015tga}, which are compatible with pentaquark states, and there are many theoretical works on deciphering their nature. On the one hand, it has been argued that these structures might not be resonances, but just kinematical effects~\cite{Meissner:2015mza,Guo:2015umn,Liu:2015fea,Mikhasenko:2015vca}. On the other hand, were it to be confirmed that the states are real resonances, the doors are still left open for different theoretical explanations to their existence: they can be described in terms of pure quark degrees of freedom~\cite{Maiani:2015vwa,Anisovich:2015cia,Mironov:2015ica,Lebed:2015tna,Esposito:2016noz}, or they might be loosely-bound meson-baryon molecular states~\cite{Chen:2015loa,Chen:2015moa,Roca:2015dva,He:2015cea,Eides:2015dtr,Lu:2016nnt}. Studies of exclusive meson electroproduction with CLAS~\cite{Mokeev:2015lda} also offer an additional opportunity for the $P_c(4450)$ structure as a combined contribution of an inner core of constituent quarks and an external meson-baryon cloud.

The goal of our work~\cite{Blin:2016dlf} has been to develop theoretical tools to help the experimental confirmation of the resonant nature of the pentaquark-like structures. In particular, we focus on the narrower of the two states, the $P_c(4450)$ of total width $\Gamma_r=39\pm5\pm19$~MeV and mass $M_r=4449.8\pm1.7\pm2.5$~MeV. We analyse the two favored spin-parity scenarios: $J_r=3/2^-$ or $J_r=5/2^+$.

It has been proposed~\cite{Blin:2016dlf,Wang:2015jsa,Karliner:2015voa,Kubarovsky:2015aaa} that in order to exclude the nonresonant scenario, photoproduction experiments would be of use: if the pentaquark-like signal is also seen in $J/\psi$ photoproduction off proton targets, then the resonant nature of the state seen by the LHCb would be confirmed, since then the explanations of kinematical effects would not apply. A further advantage of these photoproduction experiments is that the resonance peaks would be just above the threshold of $J/\psi$ photoproduction, where thus the background is very low compared to the hadronic production experiments. This should improve the possibility of observing the resonance peak, also for a relatively moderate resonance cross section.

Our previous work in Ref.~\cite{Blin:2016dlf} has helped motivating the proposal of $J/\psi$ photo- and electroproduction experiments at JLab, several of which have already been approved with high priority rating~\cite{Meziani:2016lhg}. We made predictions for the dependence of the $J/\psi$ photoproduction cross sections on the photon energy and the scattering angle, constrained by fits to the world data existing at the time. These were mainly high-energy data~\cite{Chekanov:2002xi,Aktas:2005xu}, which helped fitting the background parametrization. Concerning the low-energy data, they were extremely scarce, and only for the forward direction~\cite{Ritson:1976rj,Camerini:1975cy}. In particular, around the energy where the pentaquark state would be expected, there existed only two data points. We developed a phenomenological model for the description of $J/\psi~p$ photoproduction, fitted to the available experimental data mentioned above. This model allowed for a first prediction of the upper limit of the branching fraction for the channel $P_c(4450)\rightarrow J/\psi~p$.

Due to the renewed interest in $J/\psi$ photoproduction experiments, we are also interested in studying polarization observables, such as the asymmetry $K_{LL}$ and the spin-density matrix elements (SDMEs). In the following, we summarize our model formalism, and show the results for the abovementioned observables and the estimated upper limits.
  
\section{Reaction model}\label{model}

\begin{figure}
\begin{subfigure}[\ Pomeron exchange]{
\includegraphics[width=0.45\columnwidth]{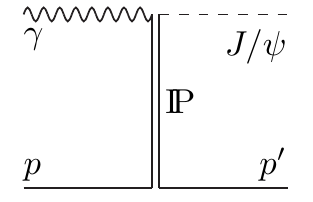}}
\label{fjpsiblob}
\end{subfigure}
\begin{subfigure}[\ Resonant contribution]{
\includegraphics[width=0.45\columnwidth]{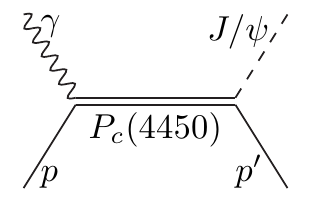}}
\label{fjpsipc}
\end{subfigure}
\caption{Model~\cite{Blin:2016dlf} contributions to the $J/\psi$ photoproduction.
The nonresonant background is modeled by an effective Pomeron exchange (a)
while the resonant contribution  of the $P_c(4450)$ in the direct channel (b) 
is modeled by a Breit-Wigner amplitude.} 
\label{Fjpsiprod}
\end{figure}

The processes contributing to $\gamma\, p \to J/\psi~p$ in our model are shown in Fig.~\ref{Fjpsiprod}. They include a non-resonant background via an effective $t$-channel Pomeron exchange, using existing high-energy data to constrain the parameters. To describe the baryon-resonance photoproduction we use the model of~\cite{Mokeev:2012vsa}, which bases on a Breit-Wigner ansatz that saturates the $s$-channel. The details are given in our previous work~\cite{Blin:2016dlf}.

It is unknown what the sizes of the couplings are, that describe the hadronic (three independent couplings $g_i$, due to the three possible helicity states of the $J/\psi$ in the final state) or electromagnetic (two independent couplings $A_i$) decays of the $P_c(4450)$ into the $J/\psi~p$ or the $\gamma~p$ channels, respectively, neither from the experiment nor from quark models. Thus, we made a first estimate: first, we set all the $g_i$ to be of the same size, $g_i=g$; then, we fixed the photocouplings $A_{1/2}$ and $A_{3/2}$ by assuming vector-meson dominance (VMD), as had already been done in Ref.~\cite{Karliner:2015voa}. By doing so, all these couplings are directly related to only one unknown parameter, the branching fraction $\mathcal{B}_{\psi p}$ of the pentaquark into the $J/\psi~p$ channel. We fitted this parameter to the experimental data, together with the parameters from the Pomeron-background piece of the amplitude.

\section{Results}\label{results}

\begin{figure*}
\begin{center}
\begin{subfigure}[\ $J_r=3/2$, $\sigma_s=0$ MeV]{
\includegraphics[width=0.8\columnwidth]{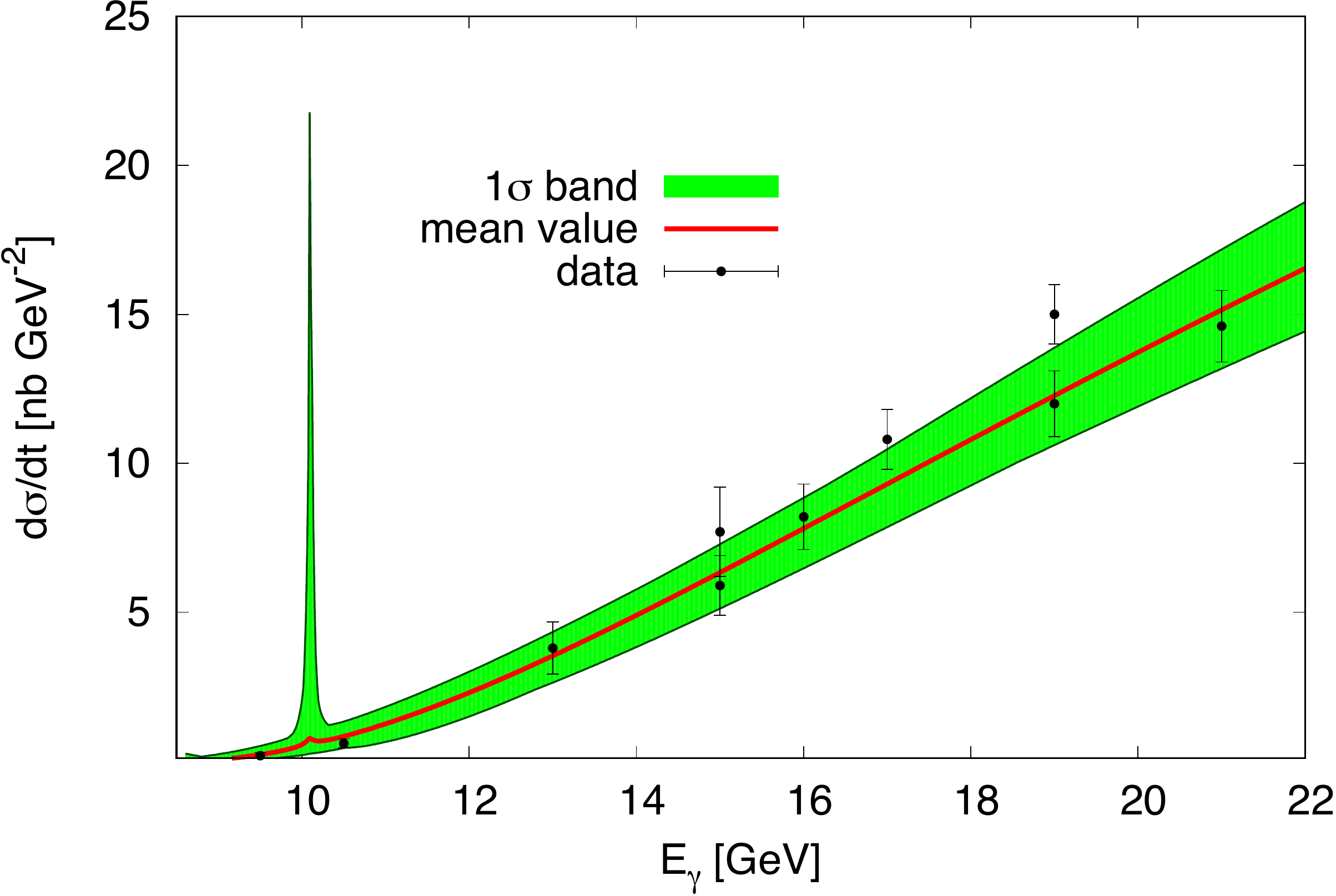}}
\end{subfigure}
\begin{subfigure}[\ $\sigma_s=60$~MeV]{
\includegraphics[width=.45\columnwidth]{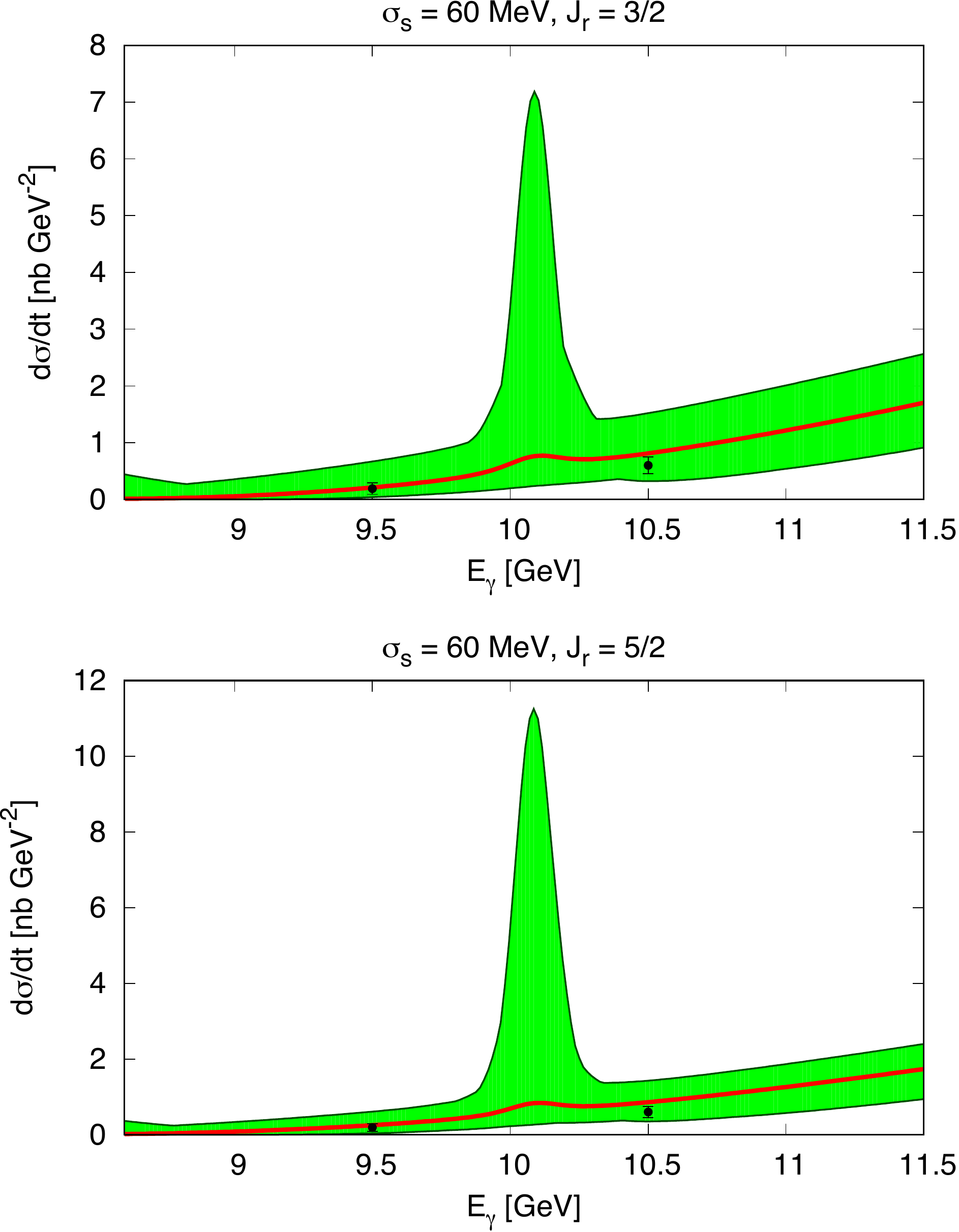}}
\label{FLow60}
\end{subfigure}\begin{subfigure}[\ $\sigma_s=120$~MeV]{
\includegraphics[width=.45\columnwidth]{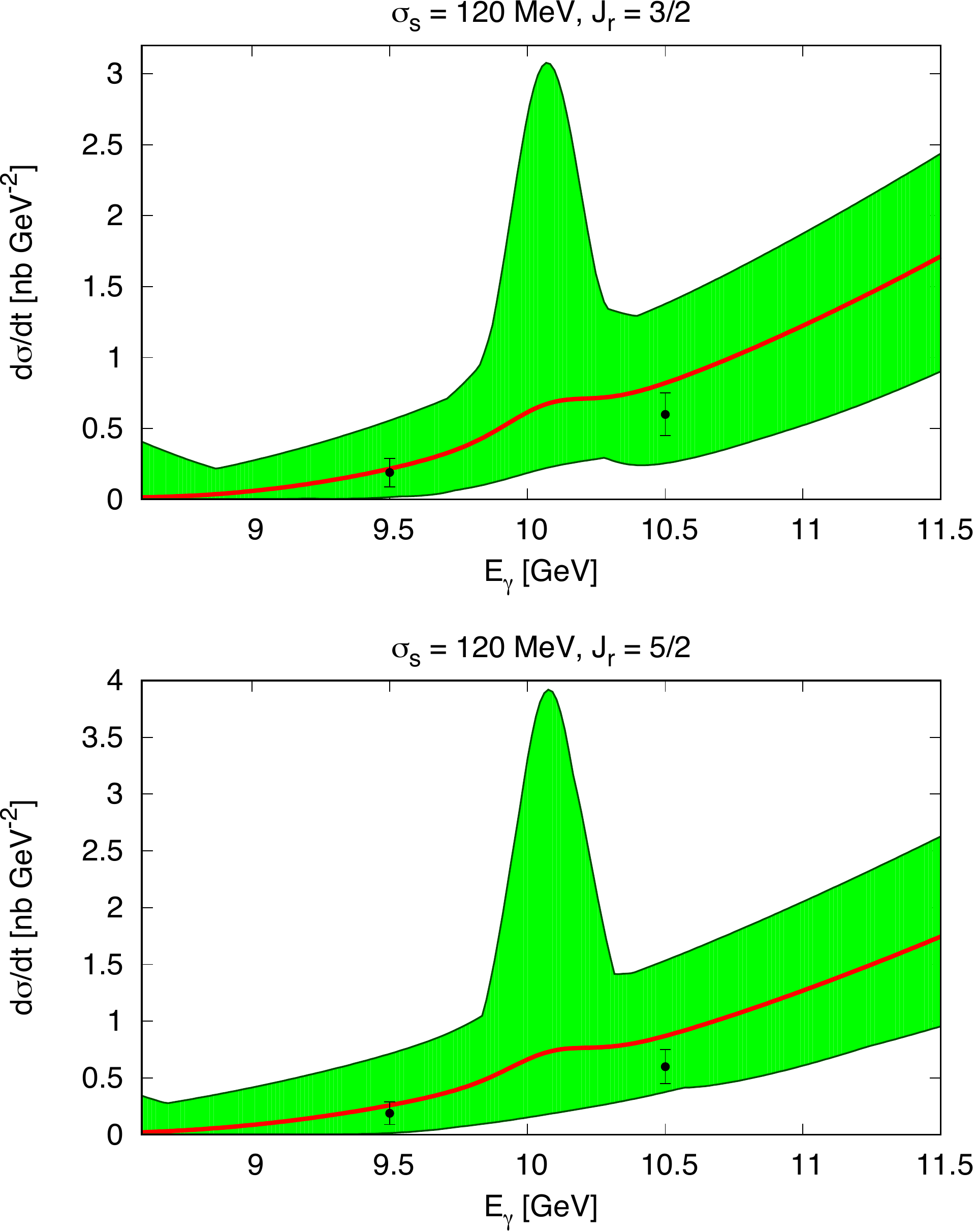}}
\label{FLow120}
\end{subfigure}%
\end{center}
\caption{Comparing data (solid circles) with the fit results at a 1$\sigma$ (68\%) C.L., 
as discussed in the text, for near-threshold differential cross section 
data~\cite{Ritson:1976rj,Camerini:1975cy} in the forward direction. 
In (a), no smearing due to the experimental resolution of the photon energy was performed, while (c) and (d) take into account a resolution of 60~MeV and 120~MeV, respectively. In (b) and (c), both the spin-3/2 (upper panel) and the spin-5/2 (lower panel) cases are shown.}
\label{FLowAll}  
\end{figure*}

Figure~\ref{FLowAll} shows the
results of the fits for different scenarios of spin assignment and smearing due to experimental resolution. For each of these scenarios, a new fit was performed. We found that the upper limit for the branching ratio $\mathcal{B}_{\psi p}$ at a 95\% C.L.\ 
ranges from $23\%$ to $30\%$ for $J_r=3/2$, 
depending on the experimental resolution, and from $8\%$ to $17\%$ for $J_r=5/2$. 
The resulting hadronic couplings, as well as the photocouplings, are summarized in Table~\ref{tabResParams} for the spin-3/2 scenario.

\begin{table*}
\caption{Upper limits at the 2$\sigma$ C.L.\ for the 
resonance parameters (as explained in Sec.~\ref{model}) obtained for the $J_r=3/2$ assignment. 
We considered three possible values for the experimental resolution $\sigma_s$. The final row, $\frac{\mathrm{d}\sigma}{\mathrm{d}t}|_{E_\gamma=E_r,t=t_\text{min}}$~(nb~GeV$^{-2})$, shows the value of the differential cross section at the resonance energy in forward direction.} 
\label{tabResParams}
\begin{center}
\begin{tabular}{c|ccc}
$J_r^P$&\multicolumn{3}{c}{$3/2^-$}\\ \hline
$\sigma_s$ (MeV) &0&60& 120\\
$\mathcal{B}_{\psi p}$ &$ \le 29\%$&$\le 30\%$ &$\le 23\%$\\
$g$~(GeV)& $ \le 2.1$&$\le 2.2$&$\le 1.9$\\ 
$A_{1/2,3/2}$~(GeV)$^{-1/2}$ &$\le 0.007$  &$\le 0.007$ &$\le 0.006$ \\
$\frac{\mathrm{d}\sigma}{\mathrm{d}t}|_{E_\gamma=E_r,t=t_\text{min}}$~(nb~GeV$^{-2}$ )
&$\le 21.8$  &$\le 7.2$ &$\le 3.1$ 
\end{tabular}
\end{center}
\end{table*}

The results allow for the existance of a structure at $10$~GeV, that can be observed in the exclusive $J/\psi~p$ photoproduction/quasiphotoproduction experiments planned at Jefferson Lab in the 12~GeV era~\cite{Meziani:2016lhg}. A fine energy-binning scan in that region might reveal this structure, since the upper limit for the branching fraction is rather large, thus seeming promising for a pronounced peak to be seen.

In view of the ongoing and future experiments at JLab, it is also interesting to study other polarization observables, such as the asymmetry between the helicities of the incoming photon and the outgoing proton $K_{LL}$ ($++$ for parallel, and $+-$ for antiparallel helicities), and the SDMEs $\rho_{ij}$:
\begin{align}
K_{LL}&=\frac{\mathrm{d}\sigma(+-)-\mathrm{d}\sigma(++)}{\mathrm{d}\sigma(++)+\mathrm{d}\sigma(+-)},\\
\rho_{mm'}&=\frac{\sum_{\mu_1\mu_2\mu_3}A_{\mu_1\mu_2\mu_3m}A^*_{\mu_1\mu_2\mu_3m'}}{\sum_{\mu_1\mu_2\mu_3\mu_4}|A_{\mu_1\mu_2\mu_3\mu_4}|^2}.
\end{align}

\begin{figure*}
\begin{center}
\begin{subfigure}[]{
\includegraphics[width=0.45\columnwidth]{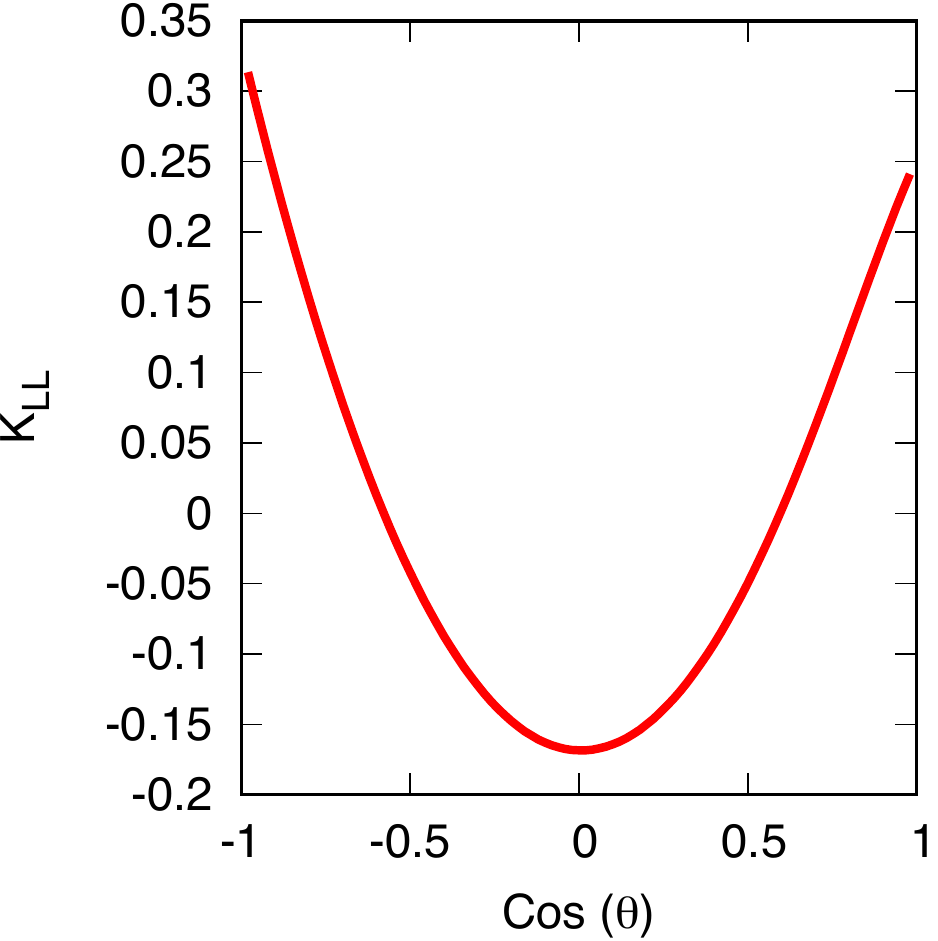}}
\end{subfigure}
\begin{subfigure}[]{
\includegraphics[width=0.45\columnwidth]{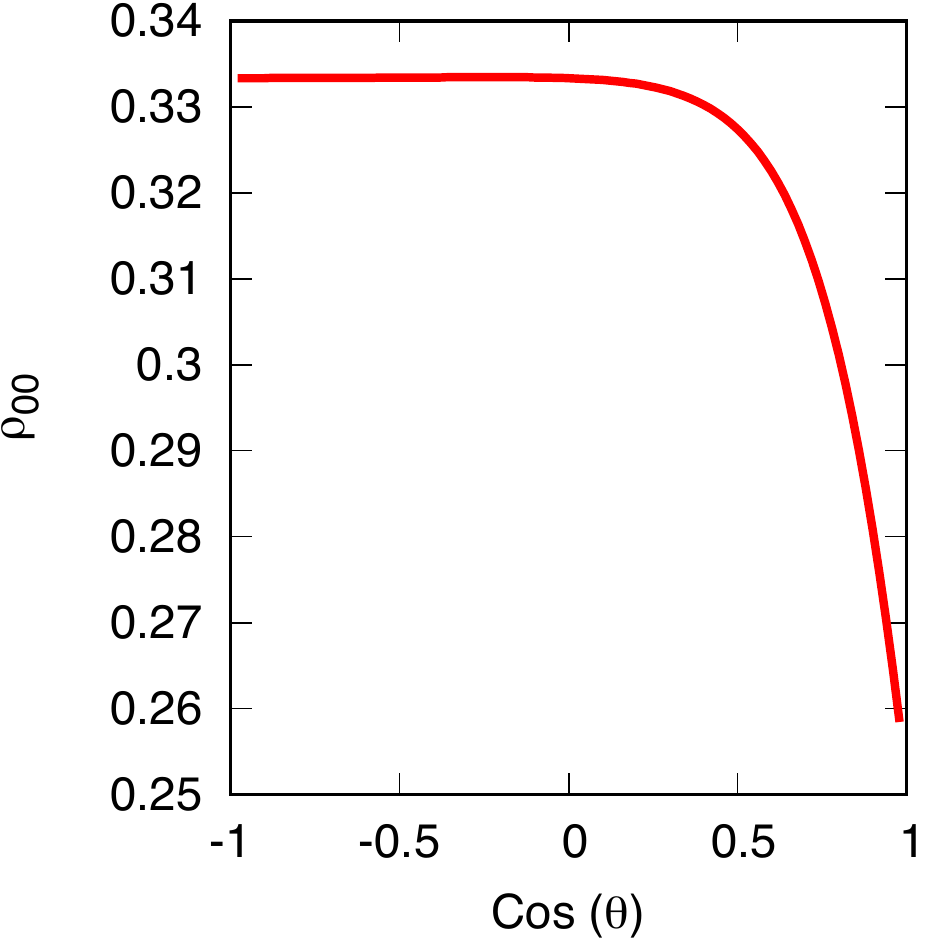}}
\end{subfigure}\\
\begin{subfigure}[]{
\includegraphics[width=0.45\columnwidth]{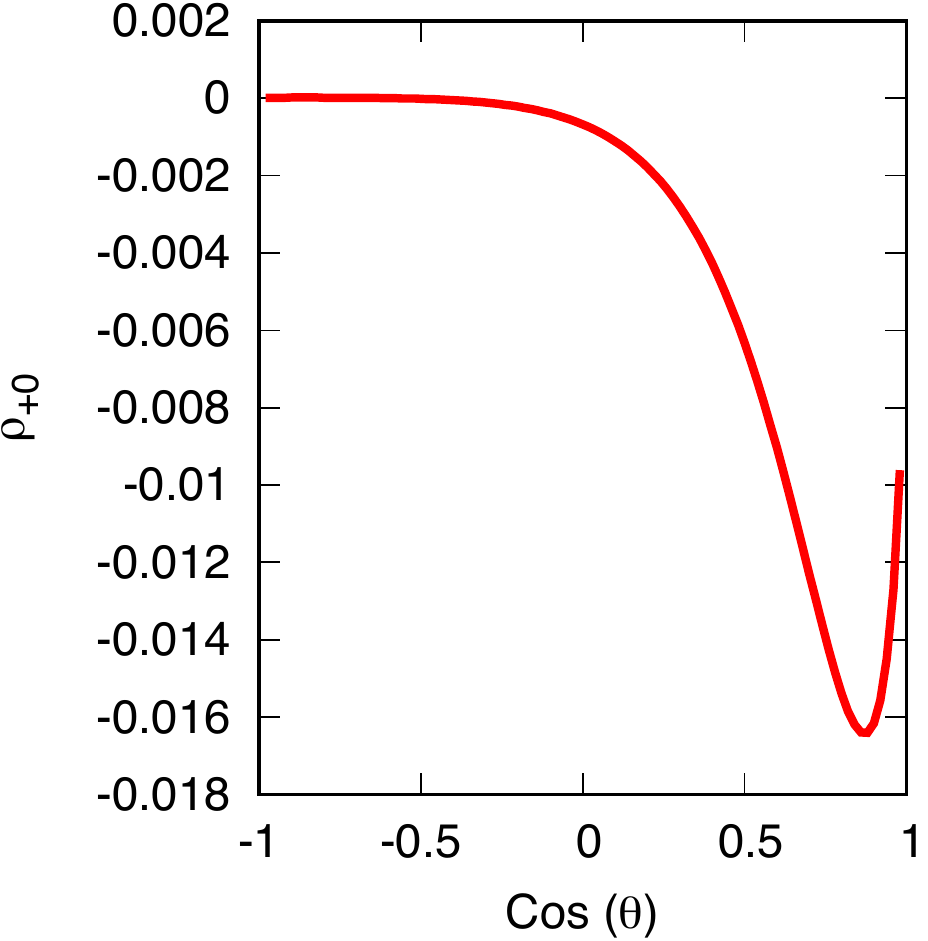}}
\end{subfigure}
\begin{subfigure}[]{
\includegraphics[width=0.45\columnwidth]{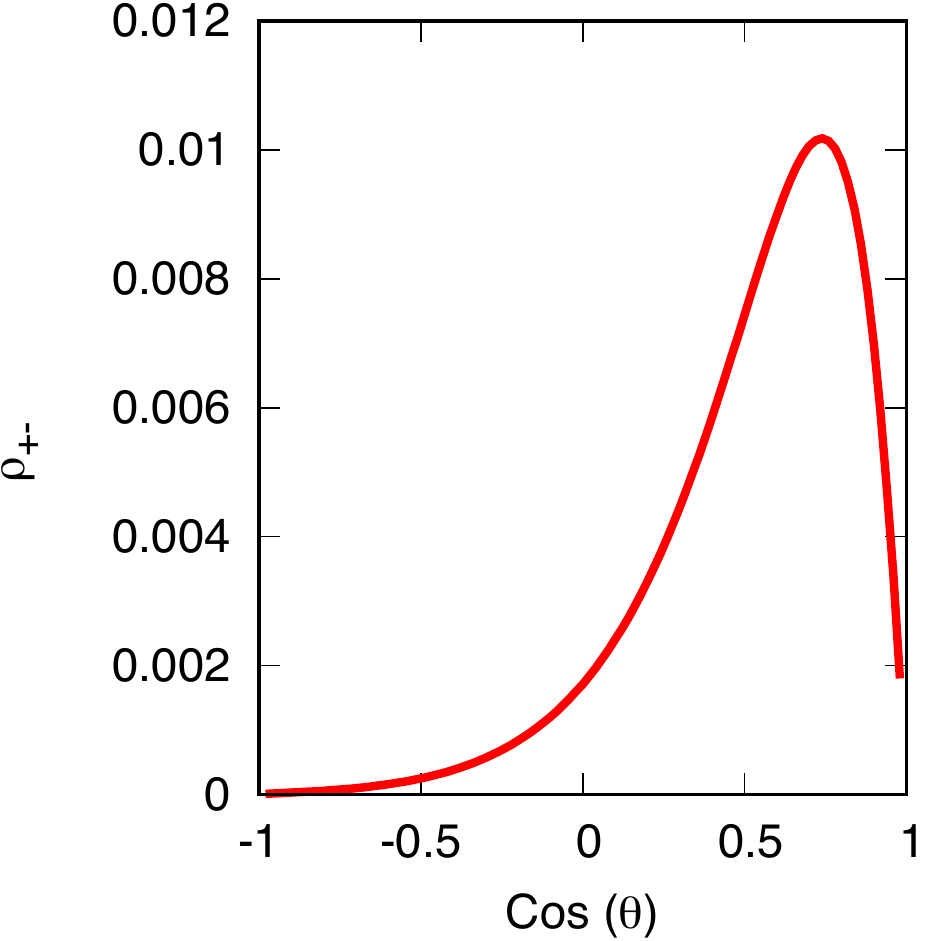}}
\end{subfigure}
\end{center}
\caption{Angular distributions of polarization observables with our fit results for the scenario of a spin-3/2 pentaquark. The observables are evaluated at the peak energy, and no smearing due to the experimental-resolution is applied. In (a), the asymmetry $K_{LL}$ is shown, and in (b), (c) and (d) the SDMEs $\rho_{00}$, $\rho_{+0}$ and $\rho_{+-}$ can be seen, respectively.}
\label{FKLL}  
\end{figure*}

In Fig.~\ref{FKLL}, our results for these observables are shown, with our best-fit parameters in the spin-3/2 case without smearing. Due to the ansatz made in our model for the background, these polarization observables are all trivially 0 for the Pomeron piece of the amplitude. They only have non-vanishing sizes for the resonant region. Currently, we have only explored the sensitivity of the aforementioned polarization asymmetry to the resonant contributions. The $P_c(4450)$ resonance generates sizable $K_{LL}$ asymmetries and pronounced angular dependencies of SDMEs at forward angles. We are currently extending our background model such as to take into account the different helicity-dependent contributions to the Pomeron amplitude~\cite{WIP:Winney2018}, thus enabling us to give predictions for these observables useful for the experiments to come.

\section{Summary}
We studied the possibility of observing the $P_c(4450)$ resonance in $J/\psi$ photoproduction in future and ongoing experiments. We used a simple two-component model 
containing an $s$-channel resonance and a $t$-channel Pomeron background.

The $P_c(4450)$ resonance signal might be observed in experiments on exclusive near-threshold $J/\psi$ photoproduction off protons planned at Jefferson Lab~\cite{Meziani:2016lhg}. We showed that the resonance peak might have escaped detection until now due to poor energy resolution, and gave predictions to observables such as differential cross sections, SDMEs and asymmetries. These can be used in the analyses of the experimental results on exclusive $J/\psi$ photoproduction aimed for the new $P_c$ state search. We also gave a first estimate for the upper limit of the branching fraction of the pentaquark into the $J/\psi~p$ channel.
 
These studies are of interest for the future experiments on the pentaquark search in exclusive $J/\psi$ photo-/electroproduction at Jefferson Lab in the 12~GeV era~\cite{Meziani:2016lhg}. It is important to mention that, if the pentaquark state is not seen in those experiments, its existence is not necessarily refuted. This fact could simply point to the small size of the photo-/electrocouplings, calling for other processes to study it better.

The code for the evaluation of all the observables studied in the present work based on the approach presented here is available on the JPAC webpage~\cite{Mathieu:2016mcy}.

\section*{Acknowledgements} 
This material is based upon work supported in part by the 
U.S.~Department of Energy, 
Office of Science, 
Office of Nuclear Physics under contracts
DE-AC05-06OR23177 and DE-FG0287ER40365, 
National Science Foundation under grants PHY-1415459 and NSF-PHY-1205019, and IU Collaborative Research Grant. 
This work was supported by the Spanish Ministerio de Econom\'ia y Competitividad (MINECO) and the European fund for regional development (EFRD) under contracts No.\ FIS2014-51948-C2-2-P and No.\ SEV-2014-0398. It has also been supported by Generalitat Valenciana under contract PROMETEOII/2014/0068 and the Deutsche Forschungsgemeinschaft DFG.


\end{document}